\documentclass{article}

\PassOptionsToPackage{numbers, compress}{natbib}

\usepackage[final]{neurips_2021}

\usepackage[utf8]{inputenc} % allow utf-8 input
\usepackage[T1]{fontenc}    % use 8-bit T1 fonts
\usepackage{hyperref}       % hyperlinks
\usepackage{url}            % simple URL typesetting
\usepackage{booktabs}       % professional-quality tables
\usepackage{amsfonts}       % blackboard math symbols
\usepackage{nicefrac}       % compact symbols for 1/2, etc.
\usepackage{microtype}      % microtypography
\usepackage{xcolor}         % colors

\title{Automated Configuration and Usage of Strategy Portfolios for Bargaining}

\author{%
    Bram M. Renting \\
    Leiden University \\
    Delft University of Technology \\
    \texttt{B.M.Renting@liacs.leidenuniv.nl} \\
    \AND %chktex 1
    Holger H. Hoos \\
    Leiden University \\
    University of British Columbia \\
    \texttt{hh@liacs.nl} \\
    \And %chktex 1
    Catholijn M. Jonker \\
    Delft University of Technology \\
    Leiden University \\
    \texttt{C.M.Jonker@tudelft.nl} \\
}
\usepackage{graphicx}
\usepackage{makecell}
\usepackage{tabularx}
\usepackage{algorithm}
\usepackage[noend]{algpseudocode}
\usepackage{tabto}
\usepackage[acronym]{glossaries}
\usepackage[disable]{todonotes}
\usepackage{mathtools}
\usepackage{amsmath}

\graphicspath{ {images/} }

\newglossaryentry{hydra}{name=HYDRA,description={a method for automatically designing algorithms to complement a portfolio}}
\newglossaryentry{autofolio}{name=AutoFolio,description={an automatically configured algorithm selector}}
\newglossaryentry{genius}{name=GENIUS,description={an automatically configured algorithm selector}}

\newacronym{anac}{ANAC}{Automated Negotiating Agents Competition}
\newacronym{smac}{SMAC}{Sequential Model-based optimization for general Algorithm Configuration}
\newacronym{sat}{SAT}{Boolean Satisfiability}
\newacronym{asp}{ASP}{Answer Set Programming}
\newacronym{csp}{CSP}{Constraint Satisfaction Problem}
\newacronym{mip}{MIP}{Mixed Integer Programming}
\newacronym{aop}{AOP}{Alternating Offers Protocol}
\newacronym{saop}{SAOP}{Stacked Alternating Offers Protocol}
\newacronym{sfm}{SFM}{Smith Frequency Model}
\newacronym{dap}{DAP}{Default Algorithm Performance}
\newacronym{smbo}{SMBO}{Sequential Model-Based Optimisation~\cite{Hutter2011SequentialOptimization}}
\newacronym{slurm}{SLURM}{Simple Linux Utility for Resource Management}
\newacronym{cov}{CoV}{Coefficient of Variance}
\DeclareMathOperator*{\argmax}{arg\,max}

\newcommand{\setting}{s}
\newcommand{\settings}{S}
\newcommand{\problem}{p}
\newcommand{\problems}{P}
\newcommand{\opponent}{o}
\newcommand{\opponents}{O}
\newcommand{\da}[1]{DA(#1)}%chktex 36

\DeclarePairedDelimiter\set\{\} %chktex 21

\newcommand{\AlgoParams}[2]{
    \foreach\x/\y [count=\i] in #1 {
        \Statex\ifnum\i=1 \textbf{#2} \fi \tabto*{1.8cm} \x \tabto*{2.5cm} \y
    }
}
\newcommand{\Input}[1]{\AlgoParams{#1}{Input}}
\newcommand{\Output}[1]{\AlgoParams{#1}{Output}}
\newcommand{\Variables}[1]{\AlgoParams{#1}{Variables}}

{}
{}

\newcommand{\f}{\mkern-2mu f\mkern-3mu}

\glsdisablehyper{}

\begin{document}

\maketitle

\begin{abstract}
    Bargaining can be used to resolve mixed-motive games in multi-agent systems. Although there is an abundance of negotiation strategies implemented in automated negotiating agents, most agents are based on single fixed strategies, while it is widely acknowledged that there is no single best-performing strategy for all negotiation settings.

In this paper, we focus on bargaining settings where opponents are repeatedly encountered, but the bargaining problems change. We introduce a novel method that automatically creates and deploys a portfolio of complementary negotiation strategies using a training set and optimise pay-off in never-before-seen bargaining settings through per-setting strategy selection. Our method relies on the following contributions. We introduce a feature representation that captures characteristics for both the opponent and the bargaining problem. We model the behaviour of an opponent during a negotiation based on its actions, which is indicative of its negotiation strategy, in order to be more effective in future encounters.

Our combination of feature-based methods generalises to new negotiation settings, as in practice, over time, it selects effective counter strategies in future encounters. Our approach is tested in an \gls{anac}-like tournament, and we show that we are capable of winning such a tournament with a \(5.6\% \) increase in pay-off compared to the runner-up agent.

\end{abstract}

\section{Introduction}
Bargaining or negotiation is a prominent method to decentrally solve mixed-motive problems through reaching mutual agreement. Problems from this area occur prominently in many real-world applications (e.g., transportation of goods using warehouse robotics, coordination of autonomous vehicles, calendar scheduling). Since the 1980s, there has been research aimed at designing computer negotiators that can replace or assist humans in negotiation. Following early contributions by~\citet{Smith1980,Sycara-Cyranski1985,Rosenschein:1986:RIC:15215,Sycara1988,jelassi1989negotiation,Klein1989,Robinson1990}, this research area has evolved considerably, and at the time of this writing, there are regular negotiation competitions (e.g., \glsfirst{anac}~\cite{BaarslagANAC2010-2015}) and standardised test-beds (e.g., \gls{genius}~\cite{Lin2014GeniusNegotiators}) that support the development of algorithmic negotiation strategies. There are now more than one hundred negotiation strategies freely available that can be used as opponents to test against --- which is important, since we know that the success of a negotiator also depends on the strategy of the opponent~\cite{Baarslag2013EvaluatingCompetition}.

The improvement of negotiation strategies over time is promising; however, we observe that the strategies almost always remain monolithic, i.e.\ single strategy with fixed behaviour for every setting. It has been observed that no single strategy is optimal for all negotiation settings~\cite{Ilany2016AlgorithmNegotiation,Lin2014GeniusNegotiators}. Therefore, a good way to further improve pay-off appears to select from a portfolio of strategies, based on the negotiation setting. This introduces the problem of algorithm selection~\cite{Rice1976TheProblem} into bargaining. An early attempt on applying algorithm selection in automated negotiation was made by~\citet{Ilany2014,Ilany2016AlgorithmNegotiation}, but they only selected a strategy based on the bargaining problem, without considering the opponent, which we know to be an important factor~\cite{Baarslag2013EvaluatingCompetition}. Furthermore, they relied on a portfolio of existing strategies to select from, which potentially limits robustness.

Our contributions in this paper are as follows: 1) we apply automated algorithm configuration techniques to not only create a single negotiation strategy, but a portfolio of complementary negotiation strategies; and 2) we introduce a procedure to learn and exploit opponent and problem characteristics during a simulated \gls{anac} tournament. The first contribution uses the approach by \citet{Renting2020AutomatedStrategies} to automatically configure negotiation strategies, which we extend by implementing \gls{hydra}~\cite{Xu2010} for portfolio construction and \gls{autofolio}~\cite{Lindauer2015AutoFolioSelector} to create a portfolio selector. Empirical results on a variety of bargaining settings show that our method beats the runner-up agent by a (comfortable) margin of \(5.6\% \).

\section{Related work}\label{sec:related}
Thanks to \gls{anac}, new negotiation strategies are developed every year and collected in the \gls{genius}~\cite{Lin2014GeniusNegotiators} test-bed, to support future research; they are categorised and empirically evaluated~\cite{Baarslag2013EvaluatingCompetition,Baarslag2014WhatStop} to provide a basis for new strategies. Most negotiation strategies contain policy parameters that influence the behaviour of the agent. To optimise the performance of the agent, the parameters need tuning. So far, tuning is mostly done manually while testing on the available opponents in the \gls{genius} test-bed. Although manual configuration is conceptually easy, it is also tedious and often leads to unsatisfactory results. Following earlier attempts at automatically configuring strategies using genetic algorithms~\cite{Matos1998,Eymann2001,Dworman1996}, or reinforcement learning~\cite{Bakker2019RLBOA,Sengupta2021AnMechanism}, a recent successful approach used a model-based algorithm procedure (\gls{smac})~\cite{Hutter2011SequentialOptimization} to automatically configure a negotiation strategy~\cite{Renting2020AutomatedStrategies}.

As there is no single best strategy for all negotiation problems~~\cite{Ilany2016AlgorithmNegotiation,Lin2014GeniusNegotiators}, we should be able to improve pay-off by exploiting differences in problem instances by selecting different strategies per negotiation setting. We see this as a variation of the algorithm selection problem~\cite{Rice1976TheProblem}. Note that algorithm selection has been successfully applied to, e.g., SAT-solving~\cite{Xu2008} and pattern recognition~\cite{Miles2009CrossSelection}. However, in the field of automated negotation, only a few attempts were made to use algorithm selection methods. \citet{Ilany2014,Ilany2016AlgorithmNegotiation} used a set of past \gls{anac} strategies and predicted which strategy would perform best on a given bargaining problem; they then entered that strategy into the negotiation session. Although they managed to improve the pay-off of the agent in this manner, they were unable to win \gls{anac}. \citet{Kawata2019MetaApproach} used a portfolio of 7 strategies that previously competed in \gls{anac}. They applied a multi-armed bandit approach to find the best performing strategy for every combination of an opponent and problem, while repeating precisely the same bargaining setting 100 times. This strategy does not generalise to new negotiation settings and problems.

\section{Preliminaries}\label{sec:preliminaries}
Agent systems that are built to negotiate contain a software-based negotiation strategy. This negotiation strategy must function according to the rules (or protocol) that is set for a negotiation setting. The protocol used in this work is the \acrlong{saop}~\cite{aydougan2017alternating}, an extension of the \acrlong{aop}~\cite{rubinstein1982perfect, osborne1994course}. A deadline of 60 seconds is used, normalised to \(t \in [0, 1]\), after which a negotiation is aborted without agreement. We refer to a bargaining problem as (\(\problem \in \problems\)), which we will negotiate between our own agent and an opponent (\(\opponent \in \opponents\)). The combination of a bargaining problem and an opponent is a bargaining setting (\(\setting \in \settings = \opponents \times \problems\)). Protocols, problems and opponents are all available through the \gls{genius}~\cite{Lin2014GeniusNegotiators} test-bed (GPL v3), which we use to benchmark our agents.

\subsection{Bargaining problem}
We negotiate over multi-issue (or multi-objective) problems that are defined according to a common standard in automated negotiation~\cite{Raiffa1985TheNegotiation,Marsa-Maestre2014FromHandbook,Baarslag2014WhatStop}. Here, an issue (\(i \in I\)) is an objective in the problem for which an agreement must be found. The set of possible solutions for an issue is denoted by \(V_i\), and the Cartesian product of all the solutions of issues in a problem forms the total outcome space (\( \prod_{i\in I} V_i = \Omega \)). An outcome is denoted by \(\omega \in \Omega \).

Preferences over the outcome space \(\Omega \) are expressed through a utility function \(u(\omega)\), such that \(u : \Omega \rightarrow [0, 1]\), where a score of 1 represents the best possible outcome. We refer to our own utility function as \(u(\omega)\) and to that of the opponent as \(u_o(\omega)\).  Negotiations are performed under incomplete information, so the utility of the opponent is predicted, which we denote as \(\hat{u}_o(\omega)\).

\subsection{Dynamic agent}\label{sec:DA}
\citet{Renting2020AutomatedStrategies} built a flexible agent and automatically configured it using \gls{smac} (described later in this section). They demonstrated that this \da{\(\theta \)} was able to win an \gls{anac}-like tournament by a significant margin. We implemented the same \da{\(\theta \)} with configuration \(\theta \in \Theta \). The full configuration space \(\Theta \) of \da{\(\theta \)} can be found in \autoref{app:configspace}. There are three types of parameters that influence the behaviour of \da{\(\theta \)}: four accepting parameters that influence when the agent accepts an offer, three bidding parameters that determine the utility to demand, and six parameters that influence searching in the solution spaces for suitable solutions.

\subsection{Automated Configuration}
Automated algorithm configuration procedures evaluate configurations of a given algorithm, observe their performance, and use this information to find better-performing configurations for a given set or distribution of problem instances. We attempt to optimise the obtained utility \(r(\theta,\setting) \in [0,1]\) by playing strategy \(\theta \) in a negotiation setting \(\setting \). As we work with a set of settings \(\settings \), we define the optimisation metric as the average utility
\begin{equation}\label{eq:performance}
    R(\theta, \settings) = \frac{1}{|\settings|} \cdot \sum_{\setting \in \settings} r(\theta, \setting),
\end{equation}

\paragraph{SMAC.}
We use the freely available general-purpose algorithm configurator \gls{smac}~\cite{Hutter2011SequentialOptimization} to automatically configure \da{\(\theta \)}, following the successful implementation of \citet{Renting2020AutomatedStrategies}. A pseudocode version of \gls{smac} can be found in \autoref{app:smac}, modified for this work. Here, \gls{smac} is used to optimise on single settings (\(\setting \in \settings \)) in a training set to significantly reduce computational expense. \gls{smac} attempts to model differences between negotiation settings through features that capture information on setting complexity. We describe these features in \autoref{sec:features}.

\subsection{Negotiation setting features}\label{sec:features}
To perform algorithm selection, we need some features that describe 1) the characteristics of the negotiation problem that we currently face, and 2) the characteristics of the current opponent. Then, given these features, algorithm selection essentially becomes an classification problem, where we map the current features to a selected negotiation algorithm from our portfolio. We also use these features to guide the model-based optimisation procedure of \gls{smac}. 

\citet{Renting2020AutomatedStrategies} created a set of features to describe a negotiation setting, which was partly based on previous work by \citet{Ilany2016AlgorithmNegotiation} and \citet{Baarslag2011}. We adopt this set of features consisting of bargaining problem features (\(X_{p}\)) and opponent features (\(X_{o}\)). An overview of the bargaining setting features we use is given in \autoref{app:features}. Opponent behaviour depends partly on the problem and is not always deterministic. We therefore calculate both the mean and covariance of the opponent features over multiple negotiation settings as opponent features for a total of 8 opponent features.

\subsection{Problem definition}\label{sec:problemdefinition}

\paragraph{Strategy portfolio creation.}
We have an agent with a dynamic strategy \da{\(\theta \)} based on configuration space \(\Theta \). Can we create a portfolio of configurations \(\boldsymbol{\theta} \subset \Theta \) using a training set of negotiation settings \(\settings \) consisting of configurations that outperform each other on specific subsets of a test set of negotiation settings \(\settings_{test}^\prime \subset \settings_{test} \) that have never been encountered before?

\paragraph{Algorithm selection.}
We have an agent with a dynamic strategy \da{\(\theta \)}, and a portfolio of configurations \(\boldsymbol{\theta} = \{\theta_1, \theta_2, \dots, \theta_n\} \), where \(\theta_1\) is the single best-performing configuration (\autoref{eq:portfoliosbs}). Can we apply an algorithm selection method \(\theta_{\setting} = AS(\boldsymbol{\theta}, \setting)\) that selects a configuration \(\theta_{\setting} \) from \(\boldsymbol{\theta} \) based on negotiation setting \(\setting \), such that \(R(AS(\boldsymbol{\theta}, \setting), \settings_{test}) > R(\theta_1, \settings_{test})\). The real goal here is to let \(R(AS(\boldsymbol{\theta}, \setting), \settings_{test})\) approach the performance of the oracle selector (\autoref{eq:oracle}) \(R(OR(\boldsymbol{\theta}, \setting), \settings_{test})\) as closely as possible.

\section{Portfolio of bargaining strategies}\label{sec:portfolio}
As a basis for algorithm selection, we need a portfolio of negotiation strategies to select from. A simple approach is to build a portfolio of negotiation strategies that already exist within the \gls{genius} environment, which is the approach used by~\citet{Ilany2016AlgorithmNegotiation}. However, for several reasons, we consider this a less than ideal approach:
\begin{enumerate}
    \item It relies on strategies that already exist, thus limiting our choices for a portfolio to strategies that have been previously implemented and are available to be re-used.
    \item The strategies might not be optimised or optimised for a different objective, resulting in a low-performance portfolio.
    \item There might be dominated strategies in the portfolio, which are outperformed in all cases by some other strategy in the portfolio, needlessly complicating the selection problem.
    \item The portfolio might not be robust. There can be negotiation instances for which all the negotiation strategies fail to achieve a decent performance, causing ``weak spots'' in our portfolio.
\end{enumerate}

\subsection{Portfolio creation}
We aim to expand upon the work of \citet{Renting2020AutomatedStrategies}, by not only automatically configuring a single negotiation strategy, but by building a portfolio of complementary strategies to better exploit differences between negotiation settings. The portfolio of strategies \(\boldsymbol{\theta}\) we create is thus a portfolio of configurations for our \da{\(\theta \)}. In our method we will therefore enforce that every strategy must add value to the portfolio:
\begin{equation}
    \forall \theta \in \boldsymbol{\theta},\ \exists \setting \in \settings,\ 
    \forall \theta^\prime \in (\boldsymbol{\theta} \setminus \theta )\ :\ r(\theta, \setting) > r(\theta^\prime, \setting)
\end{equation}

The portfolio can be viewed as a set of strategies that each specialise on a region within the bargaining setting space. Similarities in this space are found by mapping the space to the feature space. One could obtain such a portfolio by automatically configuring strategies on sets of negotiation settings that are separated in feature space by dividing the feature space either manually or using clustering techniques. However, both methods rely on human input without clear insight into the effects. The quality of the sets is disputable, as they are created based on similarities in the given feature space without regard for the performance gains thus achieved. Therefore, instead we chose to automate the portfolio creation method by using \gls{hydra}~\cite{Xu2010}, removing the requirement of human input in feature space separation.

\subsection{HYDRA}
\gls{hydra} automatically generates a portfolio given only a parameterised strategy (\autoref{sec:DA}) and a set of negotiation settings with features (\autoref{sec:features}) while using an algorithm configurator and an algorithm selector (\autoref{sec:selection}). We provide a pseudo-code description of \gls{hydra} in \autoref{app:hydra}, modified for this work.

The main idea of \gls{hydra} is to perform multiple configurator runs on an identical set of training settings, while only modifying the performance metric. Due to the modifications to the metric, the configurator produces different strategies. In \autoref{app:hydra}, the modified performance metric is computed by  ``\textit{GetModifiedPerformanceMetric}'' and formally defined as
\begin{equation}\label{eq:modifiedperformance}
    r_k(\theta, \setting) = \max \left( r(\theta, \setting),\ r(AS(\boldsymbol{\theta},\setting),\setting)  \right).
\end{equation}

The modified performance is the better of the performance of the strategy that is assessed and the performance of the strategy that is selected by the algorithm selector. By optimising using the increase of performance as compared to the current portfolio, the configurator aims to find a configuration that adds the most value to the portfolio. In the first configurator run, the default performance metric is used. The resulting configuration \(\theta_1\) is therefore a locally optimal configuration over the full set of training settings, also known as the \textit{single best strategy} in the portfolio.

\section{Strategy selection}\label{sec:selection}
The next important step in our approach is strategy selection. We now have a portfolio of strategies \(\boldsymbol{\theta}\), but still need to decide which of these strategies best fits our current problem and opponent. We therefore desire a mapping from the feature space X to a one-hot distribution over the possible strategies. This is essentially a classification problem, which we can train on examples generated from our training set. Subsequently, we hope the learned function can generalise to new bargaining problems and unknown opponents in the test set, allowing us to select the most suitable strategy from our portfolio.

\citet{Ilany2016AlgorithmNegotiation} also considered this algorithm selection problem and analysed the performance of multiple classifiers that map feature vectors to algorithms. The process of selecting a classifier and configuring the accompanying parameters can again be seen as an algorithm configuration problem. In line with the rest of this paper, we chose to automate the configuration of an algorithm selector by using \gls{autofolio}~\cite{Lindauer2015AutoFolioSelector}, leveraging the power of a broad range of algorithm selection methods and removing human bias.

\subsection{AutoFolio}
The algorithm selection system \gls{autofolio} constructs the algorithm selector. It has a range of regression and classification methods to choose from and uses \gls{smac} to determine both the selection method to use and the setting of its hyperparameters. The data \gls{autofolio} requires as input is the performance \(r(\theta,\setting)\) of every strategy (\(\theta \in \boldsymbol{\theta}\)) on every setting (\(\setting \in \settings \)) in the training set and a set of features. Its goal is to select the best-performing strategy for every negotiation setting.

\subsection{Performance measure}
We measure the algorithm selector's performance as a normalised value between a baseline and the oracle selector (\autoref{eq:oracle}) on the test set of negotiation settings. The oracle selector always makes the perfect choice for every negotiation setting and is an upper bound on the performance of an selector using the given portfolio. It is obtained by simply trying every strategy on every setting and selecting the best strategy. The \textit{single best strategy} is the strategy in the portfolio that obtains the highest performance on the full set of negotiation settings (\autoref{eq:portfoliosbs}). We refer to this strategy as \(\theta_1\), as it is the first strategy in the portfolio produced by \gls{hydra}. The performance of the single best strategy is considered to be the baseline.
\begin{align}
    OR(\boldsymbol{\theta},\setting) & \in \argmax_{\theta \in \boldsymbol{\theta}} r(\theta,\setting) \label{eq:oracle}                                                                                                                            \\
    \theta_1             & \in \argmax_{\theta \in \boldsymbol{\theta}} R(\theta, \settings) \label{eq:portfoliosbs} 
\end{align}

\section{Empirical evaluation}\label{sec:results}
\subsection{Method}
The first configurator run with the default performance metric results in the single best strategy \(\theta_1\) on the training set of negotiation settings. We aim to complement the portfolio with an additional three strategies, so we iterate through \gls{hydra} until \(k = 4\). This also allows us to analyse the performance of portfolios of size 1, 2 and 3 due to the incremental approach of \gls{hydra}. The configurations thus obtained were tested 10 times on every negotiation setting in the training set to reduce stochastic influence. Finally, the portfolio and the performance data was used along with the setting features to configure an algorithm selector using \gls{autofolio}.

\subsubsection{Input.}
Specifics on the training and test set can be found in \autoref{app:traintest}. The bargaining problem features were calculated in advance, as described in \autoref{app:features}. The opponent features can only be gathered by performing negotiations against the opponents. We gathered these features in advance for the first configurator run, by negotiating 10 times on every setting with a manually set strategy. After the first configurator run, opponent features are extracted based on negotiations with strategies that are already in the portfolio. Note that during training, we use the actual opponents utility function (\(u_o\)) to calculate the features in \autoref{app:features} to reduce estimation noise.

\subsubsection{Hardware \& budget.}
We followed \citet{Renting2020AutomatedStrategies} in terms of computational budget, in order to be able to compare results. Each run of \gls{smac} was given a 1200-hour budget, divided over 300 parallel runs. Every run was performed on a single Intel\textsuperscript{\textregistered} Xeon\textsuperscript{\textregistered} CPU core with 2 threads and 12 GBs of RAM\@. We ran \gls{autofolio} on a single dual core processor on the same computing cluster, assigned it 4 gigabytes of RAM, and provided it with a budget of 0.5 hours.

\subsection{Results}

\subsubsection{Quality of the portfolio.}\label{sec:portfolioquality}
We tested the quality of the portfolio by testing the performance (\autoref{eq:performance}) of every configuration in the portfolio on the training and testing sets of negotiation settings. The results can be found in \autoref{tab:portfolioperf}. We included ratios that indicate how often a strategy is part of the set of best strategies per setting (``Sum'' in \autoref{tab:portfolioperf}). As a final quality check, the performance of the oracle selector (\autoref{eq:oracle}) is evaluated for varying sizes of the portfolio. We present the results in \autoref{tab:ASperformance}.

\begin{table}
    \caption{Individual configuration performance on \(\settings \) and \(\settings_{test}\). The left two columns show the average utility of every individual strategy in the portfolio on the training and test set of negotiation settings. The next four columns show the fraction of the amount settings in the test set for which a single strategy belongs to a set of best performing strategies.}\label{tab:portfolioperf}
    \centering
    \fontsize{9}{10}\selectfont
    \begin{tabular}{llllllll}        
        \toprule
         & \multicolumn{2}{l}{\(R(\theta,\cdot)\)} & \multicolumn{5}{l}{\textbf{Best performing on \(\settings_{test}\) by ratio}} \\
        \cmidrule(lr){2-3}\cmidrule(lr){4-7}
        \(\theta \) & \(\settings \) & \(\settings_{test}\) & Single best & In top 2 & In top 3 & In top 4 & Sum \\
        \midrule
        \(\theta_1\) & 0.815 & 0.742 & 0.281 & 0.100 & 0.016 & 0.123 & 0.520 \\
        \(\theta_2\) & 0.788 & 0.734 & 0.167 & 0.022 & 0.020 & 0.123 & 0.333 \\
        \(\theta_3\) & 0.789 & 0.754 & 0.154 & 0.065 & 0.031 & 0.123 & 0.373 \\
        \(\theta_4\) & 0.773 & 0.721 & 0.118 & 0.058 & 0.033 & 0.123 & 0.333 \\
        \bottomrule
    \end{tabular}
\end{table}

\autoref{tab:portfolioperf} shows the results per strategy in the portfolio in the form of an individual performance over a set of settings \(R(\theta,\settings)\). It is evident that \(\theta_1\) is the single best strategy over the full training set \(\settings \). Furthermore, as every strategy is at least once the single best on individual settings (single best ratio \(> 0\)), we can conclude that every strategy contributes to the portfolio, thus satisfying the first problem statement in \autoref{sec:problemdefinition}.

Finally, \autoref{tab:ASperformance} shows us that, at every iteration of \gls{hydra}, the oracle performance of the portfolio increases on both \(\settings \) and \(\settings_{test}\). The improvement decreases on \(\settings \) as the amount of iterations increase, indicating that \gls{hydra} fills the largest ``weaknesses'' in the portfolio first.

\begin{table}
    \caption{Algorithm selector performance compared to oracle performance. The left two columns show the upper limit in average utility for various sizes of the portfolio on the training and test set of negotiation settings. The right two columns show the average utility obtained by applying the trained algorithm selector on every setting in both sets.}\label{tab:ASperformance}
    \centering
    \fontsize{9}{10}\selectfont
    \begin{tabular}{lllll}
        \toprule
         & \multicolumn{2}{l}{\(R(OR,\cdot)\)} & \multicolumn{2}{l}{\(R(AS,\cdot)\)} \\
        \cmidrule(lr){2-3}\cmidrule(lr){4-5}
        \(\boldsymbol{\theta}\) & \(\settings \) & \(\settings_{test}\) & \(\settings \) & \(\settings_{test}\) \\
        \midrule
        \( \{\theta_1\} \) & 0.815 & 0.742 & 0.815 & 0.742 \\
        \( \{\theta_1,\theta_2\} \) & 0.870 & 0.824 & 0.865 & 0.785 \\
        \( \{\theta_1,\theta_2,\theta_3\} \) & 0.875 & 0.832 & 0.869 & 0.776 \\
        \( \{\theta_1,\theta_2,\theta_3,\theta_4\} \) & 0.879 & 0.840 & 0.868 & 0.784 \\
        \bottomrule
    \end{tabular}
\end{table}

\subsubsection{Performance of the algorithm selector.}
\autoref{tab:ASperformance} shows that there is potential in the portfolio to improve utility of \da{\(\theta \)} by \(\frac{0.840 - 0.742}{0.742} \cdot 100\% \approx 13.0\% \) on the test set, if we use the oracle selector rather than \(\theta_1\). We now replace the oracle selector with the actual selector and test its performance in two ways.

\paragraph{Performance with known opponents.}
We test the absolute performance of the algorithm selector by assuming perfect knowledge of opponent features of the opponents in the test set of negotiation setting \(\settings_{test}\). The opponent features are gathered by running 10 negotiation sessions with configuration \(\theta_1\) on the test set.

We trained and tested multiple algorithm selectors on different portfolio sizes by extending the portfolio, starting with the single best strategy \(\theta_1\). We report the performance in \autoref{tab:ASperformance}. For the oracle selector \(OR\) the performance of \da{\(\theta \)} increases with the size of the portfolio. However, the performance of the algorithm selector \(AS\) plateaus on \(\settings\) after adding the fourth strategy to the portfolio. Based on the results on the training set, we conclude that the fourth strategy in the portfolio is redundant and needlessly complicates the strategy selection procedure; we therefore omitted it in the final evaluation step reported in the following.

\paragraph{Performance with unknown opponents.}
Opponent features, in contrast to the problem features, must be learned from previous encounters. Up to this point, we assumed the opponents to always be known in advance, which is not realistic. We simulate a realistic negotiation tournament where this problem occurs. The agents in \(\settings_{test}\) can also learn from their opponents, but we cannot guarantee fair learning chances due to parallelisation. To solve this, we negotiate once against all of them and then ``wiping our memory'', giving every opponent a head start.

The question arises what strategy to select at first encounters with opponents, when no opponent features are available. If strategy selection is not possible, we select the single best strategy \(\theta_1\). Opponent features are influenced by the strategy that is selected by \da{\(\theta \)}, so we simplify the feature extraction process and only gather features when strategy \(\theta_1\) is selected. This aligns with the decision to select \(\theta_1\) at first opponent encounters. The \gls{cov} of an opponent feature (\autoref{sec:features}) needs at least two samples to be meaningful, so we set a second condition to select strategy \(\theta_1\) for the first two encounters with an opponent to ``sample'' the opponent.

To obtain the results, we iterate randomly through the test settings \(\settings_{test}\) and use \da{\(AS(\boldsymbol{\theta},\setting)\)} with \(\boldsymbol{\theta} = \{\theta_1,\theta_2,\theta_3\} \) to negotiate, following the procedure as described. Additionally, we let every opponent in the test set negotiate with every other opponent in the test set on every test problem and combine the results with the results of the \da{\(\theta \)}. This procedure is repeated 10 times to reduce variance for a total of 38\,080 negotiations. The results averaged per agent show that we are capable of winning an \gls{anac}-like tournament with our \da{\(\theta \)} using the strategy selector, see \autoref{tab:anacresultsAS}. We beat the runner-up agent (MetaAgent) by \(\frac{0.788-0.752}{0.752} \cdot 100\% \approx 5.6\% \) (one-tailed t-test \(p<0.0022\)).

\begin{table}
    \caption{\gls{anac} tournament results using \da{\(AS(\boldsymbol{\theta},\setting)\)} where all scores are averaged over all bargaining settings. The goal of \gls{anac} is to obtain the highest utility. We show the top 5 agents and all the outliers for every performance measure. Here, social welfare is the summation of utility and opponent utility, Pareto distance is the smallest distance to a Pareto efficient bargaining outcome, Nash distance is the distance to the Nash bargaining solution of the problem, and agreement ratio represents the fraction of settings that resulted in an agreement. (\textbf{bold = best}, \underline{underline = worst})}\label{tab:anacresultsAS}
    \centering
    \fontsize{9}{10}\selectfont
    \begin{tabular}{lllllll}
        \toprule
        \thead{Agent} & \thead{Utility} & \thead{Opponent\\utility} & \thead{Social\\welfare} & \thead{Pareto\\distance} & \thead{Nash\\distance} & \thead{Agreement\\ratio} \\
        \midrule
        Imitator      & \underline{0.446} & \textbf{0.901} & 1.347 & 0.091 & \underline{0.428} & 0.953 \\
        GeneKing      & 0.612 & 0.783 & 1.396 & 0.065 & 0.378 & \textbf{0.994} \\
        Mamenchis     & 0.636 & 0.863 & \textbf{1.498} & \textbf{0.016} & 0.272 & 0.993 \\
        MadAgent      & 0.669 & 0.536 & \underline{1.204} & \underline{0.232} & 0.383 & \underline{0.768} \\
        AgentKN       & 0.690 & 0.757 & 1.447 & 0.065 & \textbf{0.252} & 0.934 \\
        SimpleAgent   & 0.699 & \underline{0.531} & 1.230 & 0.204 & 0.398 & 0.805 \\
        AgentF        & 0.738 & 0.679 & 1.417 & 0.076 & 0.301 & 0.941 \\
        ShahAgent     & 0.741 & 0.554 & 1.296 & 0.172 & 0.342 & 0.829 \\
        \midrule
        MetaAgent2013 & 0.746 & 0.659 & 1.405 & 0.092 & 0.284 & 0.917 \\
        MetaAgent     & 0.752 & 0.634 & 1.386 & 0.106 & 0.296 & 0.894 \\
        \midrule
        \da{\(AS(\boldsymbol{\theta},\setting)\)}  & \textbf{0.788} & 0.627 & 1.414 & 0.074 & 0.314 & 0.923 \\
        \bottomrule
    \end{tabular}
\end{table}

\begin{figure}
    \centering
    \includegraphics{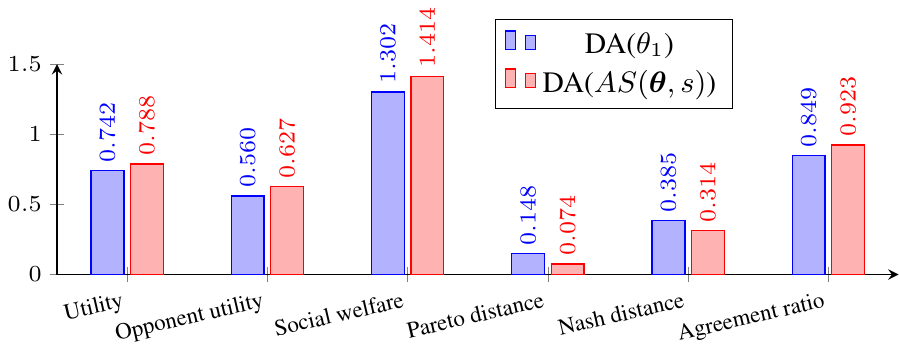}
    \caption{Comparison of two \da{\(\theta \)} strategies in an \gls{anac} tournament setting. Here, \da{\(\theta_1\)} is comparable to the agent configured by \citet{Renting2020AutomatedStrategies} and \da{\(AS(\boldsymbol{\theta},\setting)\)} represents this work. See \autoref{tab:anacresultsAS} for an explanation of the performance measures.}\label{fig:DAcomparisonanac}
\end{figure}

Finally, we compare the performances of \da{\(\theta \)} with \(\theta_1\) and with a portfolio of strategies in a realistic \gls{anac} tournament setup, see \autoref{fig:DAcomparisonanac}. Notice that our utility improved with \(\frac{0.788-0.742}{0.742} \cdot 100\% \approx 6.2\% \) by using a portfolio instead of a single fixed strategy, and that the portfolio approach also improves all other performance measures.

\section{Conclusions and future work}\label{sec:conclustion}
In previous work~\cite{Renting2020AutomatedStrategies}, automated configuration was used to obtain a single best strategy. Here, we have introduced a method to configure and use a portfolio of strategies for negotiation agents, adding a combination of \gls{hydra}, \gls{autofolio}, and a procedure to learn opponent behaviour. Our approach is fully automated and represents a significant step beyond the use of single best strategies in automated negotiation. It requires only a negotiation agent with a flexible, parameterised strategy.

We created a portfolio of 4 strategies \(\boldsymbol{\theta}\) and tested the performance of every strategy on a broad set of negotiation settings. In \autoref{tab:portfolioperf}, we showed that every configured strategy contributes to the portfolio by specialising on separate sets of negotiation settings. By adding algorithm selection to the Dynamic Agent to exploit differences between settings in a realistic tournament, we increased the performance of Dynamic Agent by \(6.2\% \) compared to the single best strategy, and won the tournament by a margin of \(5.6\% \). We note that the single best strategy is comparable to the agent configured by \citet{Renting2020AutomatedStrategies}, indicating that a portfolio-based agent provides another significant boost to negotiation pay-off.

Limitations lie in the required mutual agreement on the norms of how to conduct a negotiation. In this work, a predefined protocol is used that is supported by all used agents. Agents that do not support this protocol cannot participate in the negotiation. Another important limitation is that this method has no safeguards that check whether the currently trained agent is still performing well, and not being exploited. Finally, due to the train-then-test principle of our method, we still rely on a training set that is a decent representation of the actual application. Ethical concerns in the design of bargaining agents lie in the application of these agents in real life. Persons that have more resources to design quality bargaining strategies can gain even more resources in the process, leading to more inequality. There are risk of exploitation, unfair play, and deception due to a lack of explainability and a high level of complexity for the layman.

In future work, we intend to study the influence of Dynamic Agent's strategies on the opponent characteristics that we learn during negotiation to improve opponent learning. Secondly, strategy selection could be improved for first encounters with opponents, where currently the single best strategy is selected without regard of the setting characteristics. We want to investigate strategy selection for bargaining settings through neural networks to relax the reliance on manually designed features. Finally, it would be interesting to explore the use of reinforcement learning for training negotiation strategies instead of the algorithm configuration approach that we leveraged here.

\section*{Acknowledgement}
The authors would like to thank Thomas Moerland for his help in proof-reading this paper.

This research was (partly) funded by the \href{https://hybrid-intelligence-centre.nl}{Hybrid Intelligence Center}, a 10-year programme funded the Dutch Ministry of Education, Culture and Science through the Netherlands Organisation for Scientific Research, grant number  024.004.022 and by EU H2020 ICT48 project``Humane AI Net'' under contract \(\# \) 952026.

This research was also partially supported by TAILOR, a project funded by EU Horizon 2020 research and innovation programme under GA No 952215.

% References and End of Paper
% These lines must be placed at the end of your paper
% \bibliographystyle{ksfh_nat}
% \bibliographystyle{dinat}
\bibliographystyle{plainnat}
\newpage
\appendix

\section{Feature representations of bargaining setting}\label{app:features}
\begin{table}[H]
    \caption{Bargaining problem features (\(X_{p}\))~\cite{Ilany2016AlgorithmNegotiation}. The utility functions of the problems that are used in the paper are linear additive. An issue weight is a linear weight that is associated with an issue. The scores of every issue are multiplied by this weight and then summed to obtain the final utility. The sum of the issue weights is 1.}\label{tab:scfeatures}
    \centering
    \begin{tabular}{lll}
        \toprule
        \textbf{Description} & \textbf{Definition} \\
        \midrule
        Number of issues & \(|I|\) \\
        Average number of values per issue & \(\frac{1}{|I|} \cdot \sum\limits_{i \in I} |V_i|\) \\
        Number of possible outcomes & \(|\Omega|\) \\
        Standard deviation of issue weights & \(\sqrt{\frac{1}{|I|} \cdot \sum\limits_{i \in I} {(w_i - \frac{1}{|I|})}^2}\) \\
        Average utility of all possible outcomes & \(\frac{1}{|\Omega|} \cdot \sum\limits_{\omega \in \Omega} u(\omega) = u(\bar{\omega})\) \\
        Standard deviation utility of all possible outcomes & \(\sqrt{\frac{1}{|\Omega|} \cdot \sum\limits_{\omega \in \Omega} {(u(\omega) - u(\bar{\omega}))}^2}\) \\
        \bottomrule
    \end{tabular}
\end{table}

\begin{table}[H]
    \caption{Opponent features (\(X_{o}\))~\cite{Renting2020AutomatedStrategies}. \(x^-_o\) is the lowest offer by the opponent in their predicted utility. \(\omega^+\)/\(\omega^-\) is our best/worst possible outcome. \(\bar{x}\) is the (fictional) average offer by the opponent in their predicted utility. \(\omega_{agree}\) is the agreement.}\label{tab:oppfeatures}
    \centering
    \begin{tabular}{lll}
        \toprule
        \textbf{Description} & \textbf{Definition} \\
        \midrule
        The time it takes to reach an agreement & \(t\) \\
        Concession rate of opponent & \(\begin{cases}
            1 & \text{if } \hat{u}_{o}(x_{o}^{-})\leq \hat{u}_{o}(\omega^{+}), \\ 
            \frac{1 - \hat{u}_{o}(x_{o}^{-})}{1 - \hat{u}_{o}(\omega^{+})} & \text{otherwise.} 
        \end{cases}\) \\
        Average offer rate of opponent & \(\begin{cases}
            1 & \text{if } \hat{u}_o(\bar{x}) \leq \hat{u}_{o}(\omega^{+}),\\ 
            \frac{1 - \hat{u}_o(\bar{x})}{1 - \hat{u}_{o}(\omega^{+})} & \text{otherwise.} 
        \end{cases}\) \\
        Default strategy performance & \(\begin{cases}
            0 & \text{if } u(\omega_{agree}) \leq u(\omega^{-}),\\ 
            \frac{u(\omega_{agree}) - u(\omega^{-})}{1 - u(\omega^{-})} & \text{otherwise.} 
        \end{cases}\) \\
        \bottomrule
    \end{tabular}
\end{table}

\newpage
\section{Configuration space and configured portfolio of the Dynamic Agent}\label{app:configspace}
\begin{table}[H]
    \caption{Configuration space of \da{\(\theta \)} as set by \citet{Renting2020AutomatedStrategies}}\label{tab:configspace}
    \centering
    \begin{tabular}{llll}
        \toprule
        \thead{Description} & \thead{Symbol} & \thead{Domain} & \thead{Purpose} \\
        \midrule
        Scale factor & \(\alpha \) & \([1, 1.1]\) & Accepting \\
        Utility gap & \(\beta \) & \((0, 0.2]\) & Accepting \\ %chktex 9
        Accepting time & \(t_{acc}\) & \([0.9, 1]\) & Accepting \\
        Lower boundary & \(\gamma \) & \(\set{\text{\textit{MAX}}^W, \text{\textit{AVG}}^W}\) & Accepting \\
        Trade-off factor & \(\delta \) & \([0, 1]\) & Bidding \\
        Conceding factor & \(e\) & \((0, 2]\) & Bidding \\ %chktex 9
        Conceding goal & \(n\) & \(\set{1,2,3,4,5} \) & Bidding \\
        Population size & \(N_{p}\) & \([50, 400]\) & Searching \\
        Tournament size & \(N_{t}\) & \([1, 10]\) & Searching \\
        Evolutions & \(E\) & \([1, 5]\) & Searching \\
        Crossover rate & \(R_c\) & \([0.1, 0.5]\) & Searching \\
        Mutation rate & \(R_m\) & \([0, 0.2]\) & Searching \\
        Elitism rate & \(R_e\) & \([0, 0.2]\) & Searching \\
        \bottomrule
    \end{tabular}
\end{table}

\begin{table}[H]
    \caption{Final configurations in the portfolio. These are the final parameter settings that make up the different bargaining strategies in the portfolio.}\label{tab:portfolio}
    \centering
    % \fontsize{9}{10}\selectfont
    \resizebox{\textwidth}{!}{
    \begin{tabular}{llllllllllllll}
        \toprule
         & \multicolumn{4}{l}{\textbf{Accepting}} & \multicolumn{3}{l}{\textbf{Bidding}} & \multicolumn{6}{l}{\textbf{Searching}} \\
        \cmidrule(lr){2-5}\cmidrule(lr){6-8}\cmidrule(lr){9-14}
        \(\theta \) & \(\alpha \) & \(\beta \) & \(t_{acc}\) & \(\gamma \) & \(n_{fit}\) & \(\delta \) & \(e\) & \(N_{pop}\) & \(N_{tour}\) & \(E\) & \(R_c\) & \(R_m\) & \(R_e\) \\
        \midrule
        \(\theta_1\) & \(1.038\) & \(0.03201\) & \(0.942\) & \(AVG^W\) & \(3\) & \(0.927\) & \(0.00199\) & \(262\) & \(6\) & \(4\) & \(0.290\) & \(0.140\) & \(0.085\) \\
        \(\theta_2\) & \(1.001\) & \(0.00166\) & \(0.935\) & \(AVG^W\) & \(3\) & \(0.998\) & \(0.06232\) & \(94\) & \(2\) & \(5\) & \(0.168\) & \(0.002\) & \(0.108\) \\
        \(\theta_3\) & \(1.007\) & \(0.01970\) & \(0.912\) & \(AVG^W\) & \(4\) & \(0.917\) & \(0.01093\) & \(305\) & \(10\) & \(1\) & \(0.107\) & \(0.063\) & \(0.184\) \\
        \(\theta_4\) & \(1.056\) & \(0.00003\) & \(0.900\) & \(MAX^W\) & \(5\) & \(0.997\) & \(0.02090\) & \(139\) & \(10\) & \(4\) & \(0.463\) & \(0.176\) & \(0.101\) \\
        \bottomrule
    \end{tabular}}
\end{table}

\newpage
\section{Training and testing set of opponents and problems}\label{app:traintest}
This appendix provides an overview of the training and testing set of both opponents and bargaining problems that is used throughout this paper. A single training setting requires an agent as opponent and problem from the train set, the same is true for a test setting. 

The set of agents is provided in \autoref{tab:opponentset}. We used a total of 36 agents from the \gls{anac}. The set of \gls{anac} agents is split up in 20 training agents and 16 test agents. The set of problems is provided in \autoref{tab:scenarioset}. A total of 42 problems is used of which both sides can be played by our agent resulting in 84 playable problems. The set of bargaining problems is selected based on diversity using the features as described in \autoref{app:features} and their discount factor and reservation utility are removed. The set is split up in 56 training problems and 28 test problems.

The total amount of training settings:
\begin{equation}
    |\settings| = |O| * |P| = 20 * 56 = 1120
\end{equation}

The total amount of test settings:
\begin{equation}
    |\settings_{test}| = |O_{test}| * |P_{test}| = 16 * 28 = 448
\end{equation}

\begin{table}[H]
    \centering
    % \resizebox*{!}{0.96\textheight}{
    \begin{tabular}{l l l l}
        \toprule
        \textbf{Train/Test} & \textbf{Agent} & \textbf{ANAC} \\
        \midrule
        Test & SimpleAgent & 2017 \\
        Test & Rubick & 2017 \\
        Test & PonPokoAgent & 2017 \\
        Test & ParsCat2 & 2017 \\
        Test & ShahAgent & 2017 \\
        Test & Mosa & 2017 \\
        Test & Mamenchis & 2017 \\
        Test & MadAgent & 2017 \\
        Test & Imitator & 2017 \\
        Test & GeneKing & 2017 \\
        Test & Farma17 & 2017 \\
        Test & CaduceusDC16 & 2017 \\
        Test & AgentKN & 2017 \\
        Test & AgentF & 2017 \\
        Test & MetaAgent2013 & 2013 \\
        Test & MetaAgent & 2012 \\
        Train & ParsCat & 2016 \\
        Train & YXAgent & 2016 \\
        Train & Terra & 2016 \\
        Train & MyAgent & 2016 \\
        Train & GrandmaAgent & 2016 \\
        Train & Farma & 2016 \\
        Train & Caduceus & 2016 \\
        Train & Atlas3201 & 2016 \\
        Train & AgentHP2\_main & 2016 \\
        Train & RandomDance & 2015 \\
        Train & PokerFace & 2015 \\
        Train & PhoenixParty & 2015 \\
        Train & ParsAgent & 2015 \\
        Train & kawaii & 2015 \\
        Train & Atlas3 & 2015 \\
        Train & AgentX & 2015 \\
        Train & AgentH & 2015 \\
        Train & AgentBuyogMain & 2015 \\
        Train & Gangster & 2014 \\
        Train & DoNA & 2014 \\
        \bottomrule
    \end{tabular}
    \caption{Overview of agent set used in this work. The last column indicates in which year the agent participated in ANAC.}\label{tab:opponentset}
\end{table}

\begin{table}[H]
    \centering
    \resizebox*{\textwidth}{!}{
    \begin{tabular}{l l l l}
        \toprule
        \textbf{Train/Test} & \textbf{Profile 1} & \textbf{Profile 2} & \textbf{Comment} \\
        \midrule
        train & ItexvsCypress\_Cypress.xml & ItexvsCypress\_Itex.xml & x2 (both sides are played) \\
        train & laptop\_buyer\_utility.xml & laptop\_seller\_utility.xml & x2 (both sides are played) \\
        train & Grocery\_domain\_mary.xml & Grocery\_domain\_sam.xml & x2 (both sides are played) \\
        train & Amsterdam\_party1.xml & Amsterdam\_party2.xml & x2 (both sides are played) \\
        train & camera\_buyer\_utility.xml & camera\_seller\_utility.xml & x2 (both sides are played) \\
        train & energy\_consumer.xml & energy\_distributor.xml & x2 (both sides are played) \\
        train & EnergySmall-A-prof1.xml & EnergySmall-A-prof2.xml & x2 (both sides are played) \\
        train & Barter-A-prof1.xml & Barter-A-prof2.xml & x2 (both sides are played) \\
        train & FlightBooking-A-prof1.xml & FlightBooking-A-prof2.xml & x2 (both sides are played) \\
        train & HouseKeeping-A-prof1.xml & HouseKeeping-A-prof2.xml & x2 (both sides are played) \\
        train & MusicCollection-A-prof1.xml & MusicCollection-A-prof2.xml & x2 (both sides are played) \\
        train & Outfit-A-prof1.xml & Outfit-A-prof2.xml & x2 (both sides are played) \\
        train & RentalHouse-A-prof1.xml & RentalHouse-A-prof2.xml & x2 (both sides are played) \\
        train & Supermarket-A-prof1.xml & Supermarket-A-prof2.xml & x2 (both sides are played) \\
        train & Animal\_util1.xml & Animal\_util2.xml & x2 (both sides are played) \\
        train & DogChoosing\_util1.xml & DogChoosing\_util2.xml & x2 (both sides are played) \\
        train & Icecream\_util1.xml & Icecream\_util2.xml & x2 (both sides are played) \\
        train & Lunch\_util1.xml & Lunch\_util2.xml & x2 (both sides are played) \\
        train & Ultimatum\_util1.xml & Ultimatum\_util2.xml & x2 (both sides are played) \\
        train & DefensiveCharms\_util1.xml & DefensiveCharms\_util2.xml & x2 (both sides are played) \\
        train & SmartEnergyGrid\_util1.xml & SmartEnergyGrid\_util2.xml & x2 (both sides are played) \\
        train & DomainAce\_util1.xml & DomainAce\_util2.xml & x2 (both sides are played) \\
        train & Smart\_Grid\_util1.xml & Smart\_Grid\_util2.xml & x2 (both sides are played) \\
        train & DomainTwF\_util1.xml & DomainTwF\_util2.xml & x2 (both sides are played) \\
        train & ElectricVehicle\_profile1.xml & ElectricVehicle\_profile2.xml & x2 (both sides are played) \\
        train & PEnergy\_util1.xml & PEnergy\_util2.xml & x2 (both sides are played) \\
        train & JapanTrip\_util1.xml & JapanTrip\_util2.xml & x2 (both sides are played) \\
        train & NewDomain\_util1.xml & NewDomain\_util2.xml & x2 (both sides are played) \\
        test & England.xml & Zimbabwe.xml & x2 (both sides are played) \\
        test & travel\_chox.xml & travel\_fanny.xml & x2 (both sides are played) \\
        test & IS\_BT\_Acquisition\_BT\_prof.xml & IS\_BT\_Acquisition\_IS\_prof.xml & x2 (both sides are played) \\
        test & AirportSiteSelection-A-prof1.xml & AirportSiteSelection-A-prof2.xml & x2 (both sides are played) \\
        test & Barbecue-A-prof1.xml & Barbecue-A-prof2.xml & x2 (both sides are played) \\
        test & EnergySmall-A-prof1.xml & EnergySmall-A-prof2.xml & x2 (both sides are played) \\
        test & FiftyFifty-A-prof1.xml & FiftyFifty-A-prof2.xml & x2 (both sides are played) \\
        test & Coffee\_util1.xml & Coffee\_util2.xml & x2 (both sides are played) \\
        test & Kitchen-husband.xml & Kitchen-wife.xml & x2 (both sides are played) \\
        test & Wholesaler-prof1.xml & Wholesaler-prof2.xml & x2 (both sides are played) \\
        test & triangularFight\_util1.xml & triangularFight\_util2.xml & x2 (both sides are played) \\
        test & SmartGridDomain\_util1.xml & SmartGridDomain\_util2.xml & x2 (both sides are played) \\
        test & WindFarm\_util1.xml & WindFarm\_util2.xml & x2 (both sides are played) \\
        test & KDomain\_util1.xml & KDomain\_util2.xml & x2 (both sides are played) \\
        \bottomrule
    \end{tabular}}
    \caption{Overview of bargaining problem set used in this work}\label{tab:scenarioset}
\end{table}

\newpage
\section{SMAC}\label{app:smac}
\autoref{alg:smbo} forms the main body of \gls{smac}~\cite{Hutter2011SequentialOptimization}. The sub-procedure \textit{Intensify} is described in \autoref{alg:intensify}. We used the SMAC3 implementation of SMAC, which is released under a BSD 3-Clause License (\url{https://github.com/automl/SMAC3}).

\begin{algorithm}
    \caption{Parallel \glsfirst{smbo}}\label{alg:smbo}
    \begin{algorithmic}[1]
    \Input{{    {\(\Theta \)/Configuration space},
                {\(\settings \)/Negotiation settings},
                {\(O\)/Performance metric},
                {\(t_{opt}\)/Optimisation time budget}}}

    \Variables{{{\(R_i\)/Runhistory of pool \(i\)},
                {\(R_{{\f}ull}\)/Full runhistory of parallel pools, where \(R_{{\f}ull} = [R_1,\dots,R_m] \)},
                {\(\mathcal{M}\)/Random forest regression model},
                {\(\boldsymbol{\theta}_{new}\)/List of promising configurations}}}

    \Output{{   {\(\theta_{inc}\)/Optimised parameter configuration}}}

    \vspace{0.1cm}\hrule\vspace{0.1cm}

    \State{\([R_i, \theta_{inc}] \gets Initialise(\Theta, \settings)\)}
    \Loop{} \textbf{until} \(GetTime() > t_{opt}\)
        \State{\(R_{{\f}ull} \gets ReadParallelRunhistories()\)}
        \State{\(\mathcal{M} \gets FitModel(R_{{\f}ull})\)}
        \State{\(\boldsymbol{\theta}_{new} \gets SelectCon{\f}igurations(\mathcal{M},\theta_{inc},\Theta)\)}
        \State{\([R_i, \theta_{inc}] \gets Intensi{\f}y(\boldsymbol{\theta}_{new},\theta_{inc},R_i,\settings,O)\)}
    \EndLoop{}
    \State{\Return{} \(\theta_{inc}\)}
\end{algorithmic}
\end{algorithm}

\begin{algorithm}
    \caption{\(Intensify(\boldsymbol{\theta}_{new},\theta_{inc},R,\settings,O)\)~\cite{Hutter2011SequentialOptimization}}\label{alg:intensify}
    \begin{algorithmic}[1]
    \Input{{    {\(\boldsymbol{\theta}_{new}\)/List of promising configurations},
                {\(\theta_{inc}\)/Incumbent configuration (current best)},
                {\(R\)/Runhistory},
                {\(\settings \)/Negotiation settings},
                {\(O\)/Performance metric},
                {\(t_{int}\)/Time budget for intensify procedure}}}

    \Variables{{{\(\theta_{new}\)/Challenging configuration}}}

    \Output{{   {\(R\)/Runhistory},
                {\(\theta_{inc}\)/Incumbent configuration (current best)}}}

    \vspace{0.1cm}\hrule\vspace{0.1cm}

    \For{\(i:=1,\dots,|\boldsymbol{\theta}_{new}|\)}
        \State{\(\settings^\prime \gets \{\setting^\prime \in \settings : Count(\theta_{inc} \text{ on } \setting^\prime) \leq Count(\theta_{inc} \text{ on } \setting^{\prime\prime}), \forall \setting^{\prime\prime} \in \settings \} \)}
        \State{\(\setting \gets Random(\settings^\prime)\)}
        \State{\(R \gets ExecuteNegotiation(R,\ DA(\theta_{inc}),\ \setting)\)}
        \State{\(\theta_{new} \gets \boldsymbol{\theta}_{new}[i]\)}
        \State{\(N \gets 1\)}
        \Loop{}
            \State{\(\settings_{missing} \gets \{\setting \in \settings : Exists(\theta_{inc} \text{ on } \setting) \land \neg Exists(\theta_{new} \text{ on } \setting) \} \)}
            \State{\(\settings_{torun} \gets \) random subset of \(\settings_{missing}\) of size \(Min(N,\ |\settings_{missing}|)\)}
            \State{\textbf{for} \(\setting \in \settings_{torun}\) \textbf{do} \(R \gets ExecuteNegotiation(R,\ DA(\theta_{new}),\ \setting)\)}
            \State{\(\settings_{missing} \gets \settings_{missing} / \settings_{torun}\)}
            \State{\(\settings_{common} \gets \{\setting \in \settings : Exists(\theta_{new} \text{ on } \setting) \land Exists(\theta_{inc} \text{ on } \setting) \} \)}
            \State{\textbf{if} \(R(\theta_{new},\ \settings_{common}) < R(\theta_{inc},\ \settings_{common})\) \textbf{then break}}
            \State{\textbf{else if} \(\settings_{missing} = \emptyset \) \textbf{then} \(\theta_{inc} \gets \theta_{new}\); \textbf{break}}
            \State{\textbf{else} \(N \gets 2 * N\)}
        \EndLoop{}
        \State{\textbf{if} \((GetTime() > t_{int}) \land i \geq 2\) \textbf{then break}}
    \EndFor{}
    \State{\Return{} \([R, \theta_{inc}]\)}
\end{algorithmic}
\end{algorithm}

\newpage
\section{HYDRA}\label{app:hydra}

\begin{algorithm}
    \caption[\gls{hydra}]{\gls{hydra}~\cite{Xu2010}}\label{alg:hydra}
    \begin{algorithmic}[1]
    % \fontsize{7}{8}\selectfont
    \Input{{    {\(\Theta \)/Configuration space},
                {\(\settings \)/Training set of negotiation settings},
                {\(o\)/Performance metric}}}

    \Variables{{{\(\theta_k\)/Configuration},
                {\(\boldsymbol{\theta}\)/Portfolio of configurations},
                {\(r_k\)/Modified performance metric}}}

    \Output{{   {\(\boldsymbol{\theta}\)/Portfolio of configurations},
                {\(AS\)/Algorithm selector}}}

    \vspace{0.1cm}\hrule\vspace{0.1cm}

    % \State{\(k \gets 1\)}
    % \State{\(\theta_1 \gets RunAlgorithmCon{\f}igurator(\Theta,\settings,O)\)}
    % \State{\(TestPer{\f}ormance(\settings,\theta_1)\)}
    % \State{\(\boldsymbol{\theta} \gets \{\theta_1\} \)}
    % \State{\(AS \gets FitAlgorithmSelector(\boldsymbol{\theta}, \settings)\)}
    \State{\(\boldsymbol{\theta} \gets \emptyset \); \(r_k \gets o\)}
    % \State{\(AS \gets \) selector that always selects \(\theta_1\)}
    \For{\(k=1\); Until portfolio size is reached; \(k=k+1\)} %\textbf{until} \textit{portfolio size \(k\) is achieved}
        % \State{\(k \gets k + 1\)}
        \State{\(\theta_k \gets SMBO(\Theta,\settings,r_k)\)}\Comment{\autoref{app:smac}}
        \State{\(TestPer{\f}ormance(\settings,\theta_k)\)}
        \State{\(\boldsymbol{\theta} \gets \boldsymbol{\theta} \cup \{\theta_k\} \)}
        \State{\(AS \gets FitAlgorithmSelector(\boldsymbol{\theta}, \settings)\)}
        \State{\(r_k \gets GetModi{\f}iedPer{\f}ormanceMetric(o,AS)\)}
    \EndFor{}
    \State{\Return{} \(AS,\ \boldsymbol{\theta}\)}
\end{algorithmic}

\end{algorithm}

\end{document}